# 利用 LSTM 模型对基于 midi 格式音乐的识别与分类

**摘 要**：本研究利用人工智能神经网络对人工智能自动生成的音乐与作曲家创作的音乐进行自动识别与区分。我们首先把音频 midi 格式转化为自然语言序列进行处理,然后使用 mLSTM（multiplicative Long Short Term Memory）模型+logistic regression 回归的深度学习网络对人工智能生成音乐与作曲家创作音乐进行学习与分类，10 折交叉验证的平均准确率可达到 90%。这一工作说明人工智能生成音乐与作曲家作曲的确呈现出不同的艺术特征，这一模型也可为其他的音乐分类问题提供新的思路。

**关键词**：深度学习，自然语言处理，音乐分类

## 1 介绍

人工智能生成音乐和作曲家作曲的音乐存在很多差异，在此次比赛中，我们的主要工作是对人工智能生成的单旋律音乐进行识别。训练数据集包含 6000 个人工智能算法生成的单旋律 midi 文件以及 5742 个作曲家作曲的单旋律 midi 文件，评估数据集是由 CSMT 数据赛官方提供的 4000 个人工智能生成和作曲家作曲混合在一起的 midi 文件。本团队通过利用 mLSTM 深度神经网络模型对训练数据集进行训练，最后对评估数据集中各数据是否为人工智能作曲做出判断。

## 2 相关工作

随着深度学习的发展，除人工智能作曲之外，神经网络在各式各样的音乐分类任务上也有良好的表现。Choi K 等把音频的 log-amplitude（振幅对数）和 mel-spectrogram（频率非线性变换后的声谱）作为输入，设计了一系列 CNN 变体用于音乐的年代分类（60、70、80、90 年代），各个模型的 AUC（Area Under Curve）几乎都高于 0.7，最好的略高于 0.9[1]。Gallardo 使用 SVM 模型对音乐的各个时期进行分类：巴洛克、古典、浪漫主义、现代音乐等，把音频转换为 Humdrum 格式或 MusicXML 格式后，再进行特征提取，不同特征下的分类准确率大都在 70%-90%之间[2]。特征提取分类法用于奏鸣曲和回旋曲的曲式分类任务，准确率约为 80%[3]。神经网络还可用于音乐作品的作者分类[4]中。Micchi 使用短时傅里叶变换分析（short time Fourier transform analysis）在 6 个作曲家中确定某一作品的归属，准确率可达 70%。在音乐分类中最为常见的流派分类问题上，Costa 等把音频信号转换为声谱图，从这些时频图像中提取纹理特征，然后在分类系统中用于音乐类型的建模，最终流派分类的准确率可达 80%左右[5]。[6]中 Oramas 等分别使用音频文件、评论文字、歌曲封面对同一音乐进行多标签流派分类，结果显示使用文本信息分类的效果最好，使用音频的效果次之。

## 3 模型

音乐和语句都是与时间顺序有关的序列，音乐中各个音符的先后顺序、语句中各个字的前后关系都具有一定的意义。通过[7]和[8]可知，把自然语言模型运用在音乐样本上，也能得到类似的效果。出于以上考虑，我们把音乐样本转化为自然语言序列再进行分类，从而有效地保留音乐信息。

### 3.1 数据表示

采用[7]中的方法，按时间先后顺序把音乐片段中的信息转化为自然语言序列。这些信息包括：

（1）"n_[音高]"：音高为 0 到 127 之间的整数（包含 0 和 127）。即音高=1, 2, 3, . . . , 127.

（2）"d_[时值]_[附点数]"：时值为二全音符（breve）、全音符（whole）、二分音符（half）、四分音符（quarter）、八分音符（eighth）、十六分音符（16th）和三十二分音符（32nd）。附点数为 0、1、2、3。

（3）"v_[力度]"：力度为 4 到 128 之间 4 的倍数。即力度= 4, 8, 12, . . . , 128.

（4）"t_[速度]"：即 bpm。速度为 24 到 160 之间 4 的倍数。即速度= 24, 28, 32, . . . , 160.

（5）"."：时间步结束。每个时间步的长度和一个十六分音符的长度相同。

（6）"\n"：音乐片段结束。

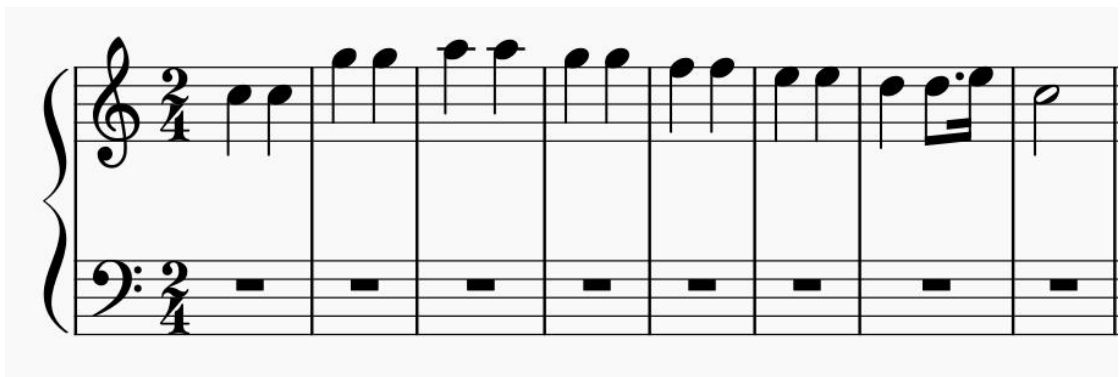

**图 1：编码小节示例**

**Fig. 1: An example bar used to be encoded by our data representation**

例如，图 1 中的这一小节乐谱（出自 12 Variations on "Ah, vous dirai-je maman",K.265/300e）转化为自

然语言序列为：

　　t_80 v_100 d_quarter_0 n_67 . . . . v_100 d_quarter_0 n_67 . . . . v_100 d_quarter_0 n_74 . . . . v_100 d_quarter_0 n_74 . . . . t_80 v_100 d_quarter_0 n_76 . . . . v_100 d_quarter_0 n_76 . . . . v_100 d_quarter_0 n_74 . . . . v_100 d_quarter_0 n_74 . . . . t_80 v_100 d_quarter_0 n_72 . . . . v_100 d_quarter_0 n_72 . . . . v_100 d_quarter_0 n_71 . . . . v_100 d_quarter_0 n_71 . . . . t_80 v_100 d_quarter_0 n_69 . . . . v_100 d_eighth_1 n_69 . . . v_100 d_16th_0 n_71 . v_100 d_half_0 n_67 . . . . . . . . t_80 .

## 3.2 mLSTM 单元与 mLSTM 层

　　mLSTM 单元（Multiplicative Long Short Term Memory Unit）与 LSTM 单元结构相似，都能记忆长期和短期的输入数据，对处理与时间顺序有关联的数据十分有效。不同之处在于 mLSTM 的权重矩阵 $W$ 依赖于每一时刻的输入，对语言进行字符级别的建模时比 LSTM 表现更好[9]。一个 mLSTM 单元的结构示意图和运算方式如下。

（1）结构示意图

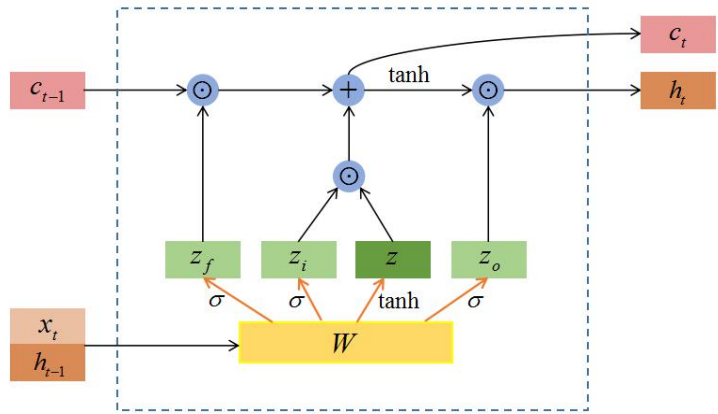

**图 2：mLSTM 单元结构示意图**

**Fig. 2: mLSTM unit structure diagram**

（2）运算方式

$$m = W_{mx}x_t \otimes W_{mh}h_{t-1}, \quad \begin{pmatrix} z_i \\ z_f \\ z_o \\ z \end{pmatrix} = \begin{pmatrix} \sigma \\ \sigma \\ \sigma \\ \tanh \end{pmatrix} W \begin{pmatrix} x_t \\ h_{t-1} \end{pmatrix}, \quad c_t = z_f \otimes c_{t-1} + z_i \otimes z$$
$$W = W_x x_t + W_h m \qquad\qquad\qquad\qquad\qquad\qquad\qquad\qquad h_t = z_o \otimes \tanh(c_t)$$

其中，$x_t$、$h_{t-1}$、$c_{t-1}$ 分别为当前时刻的输入数据、上一时刻的 mLSTM 单元隐状态（hidden state）和细胞状态（cell state），三者输入当前时刻的 LSTM 单元。$W$ 为权重矩阵。$z_f$、$z_i$、$z_o$ 分别为遗忘门（forget gate）、输入门（input gate）和输出门（forget gate），这三个门控制了当前 mLSTM 单元对输入数据的遗忘程度、处理程度和输出程度。公式中 $\otimes$ 和图中 $\odot$ 均表示矩阵的逐元素乘法，$\sigma$ 为 sigmoid 函数。经过运算，得到的 $h_t$ 和 $c_t$ 分别为当前时刻的 mLSTM 单元隐状态（hidden state）和细胞状态（cell state）。通过更新每一时刻的隐状态和细胞状态，mLSTM 可学得数据的时间变化规律。

许多个 mLSTM 单元按时间步骤排列形成一个 mLSTM 层（mLSTM layer）。 一个 mLSTM 层中的 mLSTM 单元数也称为该 mLSTM 层中神经元的个数。

### 3.3 音乐特征提取模型（mLSTM 模型）

音乐特征提取模型由一个编码层和一个 mLSTM 层构成。结构如下：

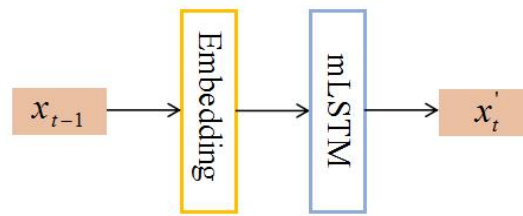

**图 3：音乐特征提取模型,$x_{t-1}$ 为上一时刻输入，$x_t^{'}$ 为所预测的当前时刻输出**

**Fig. 3: Music feature extraction model**

$x_{t-1}$ is the input of the last time step, $x_t^{'}$ is the predicted output of the current time step

使用 3.1 中的方法处理音乐样本，得到其对应的自然语言序列。取该序列某时刻的值输入 mLSTM 模型，预测下一时刻的值。mLSTM 模型 Embedding 层神经元个数为 64，mLSTM 层神经元个数为 4096，其目标函数为下一时刻的预测值与真实值的交叉熵（cross entropy loss），反向传播时更新两个层的权重，让预测值尽可能接近真实值。因为自然语言序列由音乐片段转化而来，可认为训练好的 mLSTM 模型可以根据当前时刻的音乐信息，较为准确地预测下一时刻的音乐信息。

### 3.4 分类模型

分类模型由 3.3 中已训练好的 mLSTM 模型和 logistic Regression 回归层构成，结构如下：

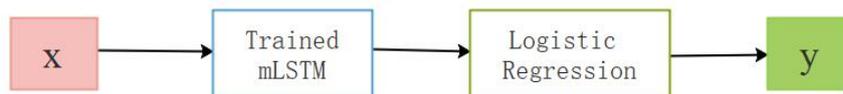

**图 4：分类模型,$x$ 为人工智能和作曲家作曲自然语言序列，$y$ 为所预测的创作类别**

**Fig. 4: Classification model**

$x$ **is the natural language sequences of AI music and composers' music, $y$ is the predicted label**

同样使用 3.1 中的方法处理人工智能和作曲家作曲的音乐样本，得到对应的自然语言序列。把序列输入训练好的 mLSTM 模型，取最终的细胞状态（cell state，一个 4096 维的向量）作为包含音乐样向量样本。

二分类逻辑斯谛回归(Logistic regression)使用极大似然估计法，根据现有数据对分类边界建立回归公式，并将分类所需预测结果输出为 0 和 1。逻辑斯谛回归将输入数据 $x_i$ 归为类别 $j$ 的概率分别为

$$p(y_i = j|x_i;\omega) = \begin{cases} \frac{1}{1+e^{w^T x_i}}, & j = 0 \\ \frac{e^{w^T x_i}}{1+e^{w^T x_i}}, & j = 1 \end{cases}$$

其中 $y_i$ 为类别标签，$\omega$ 为回归参数。

使用逻辑斯谛回归对这些向量进行分类，预测该音乐样本的时期。其损失函数为

$$L(\omega) = \sum_{i=1}^{N} [y_i(\omega^T x_i) - \log(1 + e^{w^T x_i})]$$

其中 $N$ 为样本个数。

# 4 MIDI 数据库

### 4.1 训练数据集

训练数据集由两部分开发数据集组成，一部分为 CSMT 官方给出的 6000 个人工智能算法生成的单旋律 midi 文件，另一部分为 5742 个作曲家作曲的单旋律 midi 文件。CSMT 比赛开发数据集只含人工智能算法生成音乐，包含了 6000 个单旋律 MIDI 文件，曲速在 68bpm 到 118bpm 之间，每首旋律长度为 8 小节，不包含完整的乐句结构，开发数据集中的旋律由两个音乐风格完全不同的数据库分别训练若干种不同音乐生成模型后，由算法生成。由于缺少作曲家作曲数据集，本团队通过开源网站搜寻、人工打谱、音频文件转换以及多旋律剥离主旋律等方式获得了 5742 个文件结构与官方开发数据集较一致的单旋律 midi 文件。

### 4.2 评估数据集

评估数据集为官方给出的 4000 个 MIDI 文件，除以下两点外，所有设置均与训练集相同。1）加入了一定量的作曲家的作品，其中一些为已经发表的作品，而另外一些为未发表的作品，这些作品的风格，经音乐学家鉴定，与训练算法使用的两个数据库的音乐风格分别相同。 2）评估集中存在一些由与开发集中稍微不同的算法所生成的旋律。评估集中的 MIDI 文件以下面的方式命名。

## 5 模型训练与评估

### 5.1 训练音乐特征提取模型（mLSTM 模型）

采用[7]中的方法，使用训练数据集中的音乐片段作为样本训练 mLSTM 模型。为了使训练样本多样化，对这些音乐片段进行一系列变换，包括时间变换（加速、减速），音高变换（每个音都升高或降低一个大三度）。然后采用 3.1 中的方法对样本进行编码，得到对应的自然语言序列。把这些序列拼接在一起，按 9:1 的比例随机分为训练集和测试集。再把训练集平均分为 3 个子集以便后续处理。每个训练子集中大约包括 18500 个音乐样本，测试集中大约包括 5800 个样本。

依次使用训练子集对 mLSTM 模型进行训练。在每次训练前 $h_t$ 和 $c_t$ 均初始化为 0。训练 3 轮，采用 Adam 方法进行优化，得到在测试集上的平均交叉熵损失为 0.65。

### 5.2 训练分类器

经过上一步训练的 mLSTM 模型可当做一种特殊的编码器。为了使样本数量平衡且多样化，对 MIDI 数据库中人工智能创作和作曲家作品的样本进行时间变换（加快、减慢）和音高变换（所有音上移、下移大三度）。把这些音乐样本输入 mLSTM 模型，取模型中 mLSTM 层的最终 cell state（4096 维的向量）作为编码结果，使用 logistic regression 回归进行分类。使用 10 折交叉验证（10-fold cross validation）评估模型的效果，结果如下：

**表 1：10 折交叉验证结果**

**Tab. 1: Results for 10-fold cross validation**

| 验证次数 | 准确率 |
| --- | --- |
| 1 | 99.7442% |
| 2 | 99.7442% |
| 3 | 99.5737% |
| 4 | 99.7442% |
| 5 | 99.3180% |
| 6 | 99.4881% |
| 7 | 99.6587% |
| 8 | 99.9147% |
| 9 | 99.7440% |
| 10 | 99.6587% |

在测试集上效果最好一次的混淆矩阵（confusion matrix）及分类报告（classification report）如下：

**表 2：最高准确率测试集对应的混淆矩阵**

**Tab. 2: Confusion Matrix of the highest accuracy test set**

| 实际 \ 预测 | 人工智能 | 作曲家 |
| --- | --- | --- |
| 人工智能 | 600 | 0 |
| 作曲家 | 1 | 571 |

**表 3：最高准确率测试集对应的分类报告**

**Tab. 3: Classification Report of the highest accuracy test set**

|  | 精确率（precision） | 召回率（recall） | F1 score |
|---|---|---|---|
| 人工智能 | 1.00 | 1.00 | 1.00 |
| 作曲家 | 1.00 | 1.00 | 1.00 |

由上述表格可知，mLSTM 模型+logistic regression 回归的平均准确率为 99.5 %。说明 mLSTM 是一种有效的特征提取器，可通过一个 4096 维的向量囊括一个音乐样本的大致信息。mLSTM 模型+logistic regression 回归在人工智能和作曲家作曲分类任务上有出色的表现。

5.3 模型概率

使用 5.1 中训练好的 mLSTM 模型，和 5.2 中训练好的逻辑斯蒂模型。用于预测评估数据集里面 midi 的分数，该分数代表该 midi 为人写的概率。

5.4 模型评估

mLSTM 模型+logistic regression 回归有以下几个优点：1.通过无监督学习的方法训练模型，训练时不需要大量带标签的样本，方便在标签获取成本高昂的情况下使用；2.mLSTM 能有效提取符号音乐的特征，在对音乐创作时期分类等方面有较高的准确率；3.由于 mLSTM 模型的本来目的是用于提高预测序列中后一时刻字符的准确程度，故还可用于完成音乐生成的任务。

# 6 总结与展望

本文通过 mLSTM 模型+logistic regression 回归为曲子打分（表示曲子由作曲家做的概率），来对人工智能作曲和作曲家作曲分类。通过把符号音乐转化为自然语言序列进行处理，避免了主观特征提取的困难。整个训练过程中采用无监督学习的方法，避免使用过多带标签的数据，降低了学习成本。

# 7 参考文献

classification[C]//2017 IEEE International Conference on Acoustics, Speech and Signal Processing (ICASSP). IEEE, New Orleans, LA, 2017: 2392-2396.

## 附录